\documentstyle[preprint,prl,aps,epsfig]{revtex}
\begin{document}
\preprint{LBNL-45743} 

\title{Photoproduction of top quarks in peripheral heavy ion collisions}

\author{Spencer R. Klein$^1$, Joakim Nystrand$^1$\footnote{Present address:
Division of Cosmic and Subatomic Physics, Department of Physics, Lund
University, Lund SE-22100, Sweden}, and Ramona Vogt$^{1,2}$} 
\address{$^1$Lawrence Berkeley National Laboratory, Berkeley, CA 94720 \break
$^2$Physics Department, University of California, Davis, CA 95616} 

\break 
\maketitle
\vskip -.2 in
\begin{abstract}
\vskip -.2 in 

In relativistic heavy ion collisions, top quarks can be produced by
photon-gluon fusion when a photon from the Weizs\"acker-Williams
virtual photon field of one nucleus interacts with a gluon in the
other nucleus.  Photoproduction with heavy ions at the Large Hadron
Collider (LHC) will be the first accessible non-hadronic top
production channel.  We calculate the $t \overline t$ photoproduction
cross sections, pair mass and top quark rapidity distributions in
peripheral lead-lead and oxygen-oxygen collisions.  
The cross sections are sensitive to
the top quark charge and the large-$Q^2$ gluon distribution in the
nucleus.  We find a cross section of 15 pb in oxygen-oxygen 
collisions, leading to 210 pairs in a one month ($10^6$ s) LHC run.
In $pA$ collisions, the rate is higher, 1100 pairs per month for $p$O.
A comparison of the $AA$ and $pA$ data might allow for a study of gluon
shadowing at high $Q^2$.

\end{abstract}
\pacs{14.65.Ha, 25.20.Lj, 13.60.-r}
\narrowtext

Due to their large charges, relativistic heavy ions carry strong
electromagnetic fields which may be treated as virtual photon fields.
In a relativistic ion collider, these fields interact with target
nuclei in the opposing beam, resulting in high luminosities for
photonuclear interactions.  Because of the large Lorentz boosts, high
photon-nucleon center of mass energies are reached. Previous studies
have considered photoproduction of charm\cite{bbar} and
bottom\cite{bbar2} quarks as well as nuclear breakup\cite{baurrev} and vector
meson production \cite{usPRC}.  These calculations all considered peripheral
collisions, with impact parameter, $b$, greater than twice the nuclear radius 
$R_A$ so that the two nuclei do not interact
hadronically.  Peripheral collisions will be studied
experimentally at RHIC\cite{usRHIC} and the LHC\cite{FELIX}, now under
construction at CERN.

Here, we consider the production of top quarks via photon-gluon
fusion, paralleling previous calculations of photoproduction in heavy ion
collisions \cite{bbar,bbar2}.  We calculate top production from the heaviest,
Pb, and lightest, O, ions planned for the LHC.  In these collisions, a
$t \overline t$ is produced in the reaction 
$\gamma(k) + A(P) \rightarrow t(p_1) + \overline t(p_2) + X$ 
where $k$ is the four momentum of the photon emitted from the
virtual photon field of the projectile nucleus, $P$ is the four momentum of 
the interacting nucleon in target nucleus 
$A$, and $p_1$ and $p_2$ are the four momenta of the
produced $t$ and $ \overline t$ quarks.  Note that a phton from the target can
also interact with a nucleon in the projectile.
We work in the center of mass (lab)
frame.  The photons are almost real, with 
$|q^2| < (\hbar c/R_A)^2$.  The slight virtuality is neglected.

On the parton level, the
photon-gluon fusion reaction is $\gamma(k) + g(x_2 P) 
\rightarrow t(p_1) + \overline t(p_2)$ where $x_2$ is the 
fraction of the target momentum carried by the gluon.
To lowest order (LO), the $t \overline t$ production
cross section is\cite{joneswyld}
\begin{equation}
s^2 \frac{d^2 \sigma}{dt_1 du_1} = \pi \alpha_s(Q^2) \alpha e_t^2 \left(
\frac{t_1}{u_1} + \frac{u_1}{t_1} + \frac{4m^2 s}{t_1 u_1} \left[ 1 - \frac{m^2
s}{t_1 u_1} \right] \right) \delta(s + t_1 + u_1)
\label{partonic}
\end{equation}
where the partonic invariants $s$, $t_1$, and $u_1$ are defined as $s
= (k + x_2 P)^2$, $t_1 = (k - p_1)^2 - m^2 = (x_2 P - p_2)^2 - m^2$, and 
$u_1 = (x_2 P - p_1)^2 - m^2 = (k - p_2)^2 - m^2$ 
with top quark mass $m = 175$ GeV.  
In this case, $s = 4k \gamma x_2m_p$ where
$\gamma$ is the Lorentz boost of a single beam and $m_p$ is the proton
mass.  Here, $e_t=2/3$ is the expected top
quark charge, $\alpha=e^2/\hbar c$ is the electromagnetic coupling
constant, and $\alpha_s(Q^2) \approx 0.11$ is the one-loop
strong coupling constant evaluated at scale $Q^2 = m^2 + p_T^2$ where $p_T$
is the transverse momentum of the produced top quark.  
The large mass of the top
quark prevents toponium production, allowing the $t$ and $\overline t$ to be
treated as free quarks, even near threshold \cite{toponium}.  

The hadronic top
production cross section is obtained by integrating
Eq.~(\ref{partonic}) over $x_2$ and $k$, since virtual photons are emitted by 
the nucleus in a continua of four-momenta,
\begin{equation}
S^2\frac{d^2\sigma_{\gamma A \rightarrow t \overline t X}}{dT_1 dU_1} =
2 \int_{k_{\rm min}}^\infty dk {dN\over dk} \int_{x_{2_{\rm min}}}^1 
\frac{dx_2}{x_2} g(x_2,Q^2)  s^2 
\frac{d^2 \sigma}{dt_1 du_1} \, \, ,
\label{main}
\end{equation}
where $dN/dk$ is the photon flux. The factor of two in Eq.~(\ref{main})
arises because both
nuclei emit photons and thus serve as targets.  
The incoherence of top production guarantees that there
is no interference between the two production sources\cite{usinterf}.
The hadronic invariants can be defined for a given photon four momuntum $k$. 
If the top quark is detected, the 
invariants are $S = (k + P)^2$, $T_1 = (P - p_1)^2 - m^2$, 
and $U_1 = (k - p_1)^2 - m^2$ \cite{SvN}.
The partonic and hadronic invariants are related by $s = x_2S$, $t_1 = U_1$,
and $u_1 = x_2 T_1$.  Four-momentum conservation at leading order gives
%$s + t_1 + u_1 = x_2 S + U_1 + x_2 T_1 = 0$ so that 
$x_{2_{\rm min}} = -U_1/(S
+ T_1)$.  In addition to the total cross sections, we also present the top
quark rapidity distribution and the $t \overline t$ invariant mass
distribution.  We define the $t$ rapidity as $y$ and the $\overline t$ 
rapidity as
$y_2$.  The top quark rapidity is related to the invariant $T_1$ by 
$T_1 = - \sqrt{S} Q e^{-y}$.  The invariant mass of the pair can be determined
if both the $t$ and $\overline t$ are detected.  The invariant mass
is $s = M^2 = 
2Q^2 (1 + \cosh(y - y_2))$. The minimum photon momentum necessary to produce
a $t \overline t$ pair is $k_{\rm min} = M^2/(4\gamma x_2 m_p)$.
%Energy conservation determines $k_{\rm min}=m^2/(2\gamma^2-1)x_2 m_p$.  

We use the MRST LO gluon distribution $xg(x,Q^2)$\cite{pdf} which is 
considerably softer than the older, flat
parameterizations such as $xg(x) \propto (1-x)^n$\cite{frstr} 
used in earlier photoproduction predictions.  Calculations with $n= 5$ result
in cross sections $2 - 3$ times higher than those with the MRST LO gluon
distribution.  Shadowing, the modification of the gluon
distribution in nuclei, is included via a parameterization based
on a fit to data\cite{eskola}.  There are two caveats regarding
shadowing.  First, the large $Q^2$ of top production is outside the
upper limit of the parameterization, $Q_{\rm max}^2 = 10^4$ GeV$^2$.  However,
the shadowing parameterization continues to evolve beyond this point 
with changes on the percent level between $Q^2 = 10^4$ GeV$^2$ and $6.4 \times
10^5$ GeV$^2$.  Secondly, the
probable impact parameter dependence of the shadowing \cite{ekkv4} has
been neglected.  

The photon flux is given by the Weizs\"acker-Williams
formulae.  The flux is a function of the distance
from the nucleus, $r$,
\begin{equation}
{d^3N\over dkd^2r} = {Z^2\alpha w^2\over \pi^2kr^2} \left[ K_1^2(w) + {1\over
\gamma^2} K_0^2(w) \right] \, \, ,
\label{wwr}
\end{equation}
where $w=kr/\gamma$ and $K_0(w)$ and $K_1(w)$ are modified Bessel
functions.  The photon flux is cut off at an energy determined by the
size of the nucleus.  In the rest frame of the target nucleus, the
cutoff is boosted to $(2\gamma^2-1)\hbar c/R_A$, or 500 TeV for lead
and 1800 TeV for oxygen.  The $t \overline t$ production cross section, 
$\gamma p \rightarrow t \overline t$ \cite{frstr}, for a photon with 
the cutoff energy of
oxygen is a factor of 7.2 times larger than the cross section at the
cutoff energy of lead.  Thus the difference in the energy cutoffs of
lead and oxygen is significant.

The total photon flux striking the target nucleus is the integral of
Eq.~(\ref{wwr}) over the transverse area of the target at all impact
parameters subject to the constraint that the two nuclei do not
interact hadronically\cite{bbar,usPRC}.  The numerical result
agrees to within 15\% of the analytical result,
\begin{equation}
{dN\over dk} = {2Z^2 \alpha \over\pi k} \left[ w_RK_0(w_R)K_1(w_R)-
{w_R^2\over 2} \big(K_1^2(w_R)-K_0^2(w_R)\big) \right] \, \, ,
\label{flux}
\end{equation}
the integral of Eq.~(\ref{wwr}) for $r > 2R_A \approx 2.4 A^{1/3}$
where $w_R=2kR_A/\gamma$. The 15\% difference can serve as a
conservative estimate of the uncertainty on the photon flux; this can
be checked using other photoproduction and two-photon interactions.

Although our $t \overline t$ calculation is at leading order, the
large mass of the top quark ensures faster convergence of the
perturbative expansion than for the lighter charm and bottom quarks
where the next-to-leading order corrections lead to factors of $2-3$
enhancements over the LO cross section.  The next-to-leading order $t
\overline t$ cross sections are only $\approx 40$\% larger than the LO
cross sections in $pp$ interactions \cite{MNR}.  We assume the NLO
enhancement to be similar for photoproduction.

The expected center of mass
energies and average luminosities for O+O and Pb+Pb collisions at the LHC
are given in Table~\ref{constant}. 
The numbers are taken from the latest LHC machine
studies on LHC ion operation \cite{brandt}. The calculation of the average
luminosities assumes two experiments, {\it e.g.}\
ALICE and CMS, and a bunch spacing of 125~ns.

The total cross sections for $t\overline t$ pair production with lead
and oxygen beams are 550 pb and 15 pb.  As Table~\ref{constant} 
shows, a $10^6$ s (one
month) heavy ion run at design luminosity will produce
0.2 and 210 pairs respectively.  The most significant factor in the
rate difference comes from the luminosity, $\sim 3 \times 10^4$ 
times higher for oxygen.

Shadowing plays little role in the total cross sections since at $y =
0$ and $p_T = 0$, $\langle x_2 \rangle \approx 0.06$ for lead and 0.05
for oxygen.  These values are in a region where shadowing effects are small.
In addition, at the large $Q^2$ required for top
production, shadowing effects are further reduced by $Q^2$ evolution
\cite{eskola}.  The uncertainty in the large-$Q^2$ shadowing is
important since no data are available.  However, because of the 
large luminosity gain, lighter mass ions
are better for studies of nuclear gluon distributions, reducing any possible
effect still further.

The top quark rapidity distributions, obtained from
Eq.~(\ref{main}),  are shown in Fig.~1.  The calculations 
assume that the photon is in the field of the 
nucleus coming from positive rapidity so that $y<0$
corresponds to $k<\gamma x_2m_p$ in the center of mass (lab) frame.  If the
photon is emitted by the target instead of the projectile, the resulting 
top quark rapidity 
distribution would then be the mirror image of the distribution in Fig.~1
around $y=0$. (Note that this mirror distribution is 
equivalent to detecting the $\overline t$ instead of the $t$.  
Thus the $t$ and $\overline t$ distributions are not
symmetric around $y=0$.)  The total top quark rapidity distribution is
the sum of the curve in Fig.~1 with its mirror image when both nuclei emit
photons. 
%Since either nucleus can emit a photon, the
%total $d\sigma/dy$ is the sum of this curve and its mirror image around $y=0$.
Roughly half the production is within $|y|<1$, in the central ALICE acceptance 
and almost all $t \overline t$ production falls within the CMS acceptance, $|y|
< 2.4$.
The $t\overline t$ pair invariant mass distributions, $d\sigma/dM$, are 
shown in Fig.~2. The larger photon energy of 
oxygen results in a broader pair mass distribution. 

The only previous study of $t\overline t$ production via photon-gluon
fusion in heavy ion collisions\cite{schneider} used a very different
photon flux.  The flux was integrated over all $r$, including the
nuclear interior, modelling the nucleus as a homogenously charged
sphere.  This approach is flawed because the photon flux inside the
nucleus is poorly defined.  Ref.~\cite{fflux2} shows that the photon
flux varies by orders of magnitude inside the nucleus, depending on
the chosen nuclear form factor.  More importantly, the calculation is
incorrect because the photon flux inside the nucleus is much higher than
that outside the nucleus.  The authors apply a correction factor of $1/2$
but this factor is inappropriately large.  Their photon flux (and corresponding
cross section) for Pb+Pb collisions is a factor of 10 greater than ours.

Top quark pairs can be observed via their decays, predominantly
$t\overline t\rightarrow W^+ b W^-\overline b$, where the $W$ decays
to $\ell\nu$ or $q\overline q'$.  The major background to these
channels is likely to be hadroproduction of top in grazing peripheral
collisons with slightly smaller impact parameter.  The two reactions
can be separated based on the presence of rapidity gaps in the
collision, the breakup of the colliding nuclei, and a small
multiplicity difference.  Because the photon is colorless, even if the
nucleus breaks up, photoproduction events should also have a rapidity
gap between the photon-emitting nucleus and the $t\overline t$ system.
The average multiplicity in $pp$ collisions at the LHC is expected to
be about 45, slightly higher than the roughly 35 particles expected in
a $\gamma p$ collision at typical top-production energies\cite{caso}.
The photoproduction multiplicity should be further reduced since much
of the photon energy is needed to produce the $t \overline t$ pair,
leaving less energy for produce additional particle production.   

For lead, nuclear breakup can be a complication.  A lead nucleus may be
excited into a giant dipole resonance with probability 
$\approx 35\%\cdot (b/2R_A)^2$.  When the resonance decays,
the nucleus will emit one or more neutrons\cite{hencken}.  To a good
approximation, the excitation and the photon-gluon fusion occur
independently.  Thus the two processes factorize and can be studied
separately \cite{na50fission}.  For oxygen, the excitation probability
is very small, so one of the interacting ions will almost always
remain intact.

The photoproduction cross sections are large compared with the
corresponding $pp\rightarrow t\overline tX$ cross sections.  At lowest
order, $\sigma(pp\rightarrow t\overline tX)$ = 41 pb at $\sqrt{s} =
5.5$ TeV, 7\% of the lead photoproduction cross section, while
$\sigma(pp\rightarrow t\overline tX)$ = 80 pb at $\sqrt{s} = 7$ TeV,
5.3 times the corresponding oxygen photoproduction cross section.
Thus, only a moderate hadronic rejection factor is required.

Other backgrounds should be small.  Hadronic single or double
diffractive production without accompanying colored interactions,
$AA\rightarrow AA\overline t t X$ occurs only in a narrow range of
impact parameters.  Diffractive production is also suppressed by the
$1/M^4$ final-state mass dependence.  Any single diffractive top
production will be at larger rapidities than photoproduction which is
more central.  We have calculated $\sigma(\gamma \gamma \rightarrow t
\overline t)$ and found it to be negligible.  Backgrounds from other
photoproduction channels should be significantly smaller than at
hadron colliders because $\sigma(\gamma p \rightarrow Q \overline Q
X)/\sigma(\gamma p \rightarrow X)$ is much larger than $\sigma(p p
\rightarrow Q \overline Q X)/\sigma(p p \rightarrow X)$ at comparable
energies.

We believe that these 
criteria should allow for at least statistical separation of
events containing photoproduction of top in oxygen on oxygen
collisions.

The top photoproduction cross sections are also measurable in $pA$
collisions, as may be possible at the LHC\cite{FELIX,brandt}. Because
the proton and the ion must have the same magnetic rigidity and $Z=A$
for the proton, these collisions are at somewhat higher per nucleon
energies than the corresponding $AA$ collisions.  The cost is that the
center of mass is no longer at rest in the lab.  Our $pA$ results are
calculated in the equal speed system so that the $\gamma$
is that of the equal speed system.  In this case, the photon flux is calculated
using the analytical expression in Eq.~(\ref{flux}) with $w_R = (r_p + 
R_A)k/\gamma$ and $r_p = 0.6$ fm is the proton radius. 

The expected
nucleon-nucleon center-of-mass energies and luminosities for $pA$ collisions
are shown in
Table~\ref{pa}.  No official LHC $pA$ luminosities are yet available
\cite{brandt2}. The values in Table~\ref{pa} are based on estimates in
Ref.~\cite{morsch}, which were obtained assuming a 125~ns bunch
spacing and one experiment.  

The $pA$ collisions would allow a measurement of the gluon structure
function in free protons.  A comparison between photoproduction in
$pA$ and $AA$ collisions at the same energy would provide a
straightforward measurement of nuclear gluon shadowing at $Q^2$ values
far above those currently available.  The rates are much higher at the
maximum $pA$ energies, however, since the cross sections per nucleon
are larger than in $AA$ collisions. The center-of-mass energy is
40-60\% higher, leading to a larger boost in the target rest frame,
increasing the photon cutoff energy to 1200~TeV for lead, more than a
factor of two greater than in Pb+Pb, and 3600~TeV for oxygen.  In
addition, the average photon flux is higher in $pA$ because single
protons can approach the photon-emitting nucleus more closely than a
proton in the center of another nucleus.  The estimated luminosities
for $pA$ interactions lead to 40 pairs in $p$Pb and 1100 pairs in $p$O
over a $10^6$ s run, as shown in Table~\ref{pa}.

Although the top quarks are produced inside the target nucleus,
because of their very high boost with respect to the target, they
decay well outside the nucleus.  For a standard model top width of
$\Gamma_t=$1.5 GeV\cite{toprev} the top quarks typically travel over
100 fm in the nuclear rest frame before decaying.  Thus the
$t\overline t$ pair acts as a dipole with separation $\hbar c/m \sim
10^{-3}$ fm, resulting in a small interaction cross section.
Therefore, aside from debris from the target nucleon, the target
nucleus will be relatively undisturbed, leaving the $t\overline t$
pair with relatively few accompanying particles.

In conclusion, top quark pairs will be produced via photon-gluon
fusion in heavy ion collisions at the LHC.  Heavy ion photoproduction
will be the first accessible non-hadronic production channel for
$t\overline t$ pairs.  A $10^6$ s O+O run will produce 210 $t\overline
t$ pairs while $pA$ runs will result in higher rates, up to 1100 pairs
in $p$O collisions.  These events could at least be statistically
identifiable on the basis of accompanying rapidity gaps, the presence of
an intact nucleus, and a slightly smaller multiplicity.  The data should
allow for measurements of the top charge and mass.  In addition, the
nuclear gluon distribution may be measurable at large $Q^2$.

This work was supported in part by the Division of Nuclear Physics of the 
Office of High Energy and Nuclear Physics of the U. S. Department of Energy
under Contract No. DE-AC-03-76SF00098.

\begin{table}
\begin{tabular}{lcccc}
Ion Species & $\sqrt{S_{NN}}$  & $AA$ Luminosity & $\sigma(AA)$ & Rate \\
        & (TeV) & (pb$^{-1}$s$^{-1}$) & (pb) & (per month) \\ \hline
Oxygen  & 7.0  & $1.4\times10^{-5}$   & 15  & 210  \\
Lead    & 5.5  & $4.2\times10^{-10}$  & 550 & 0.2  \\
\end{tabular}
\caption[]{Center of mass energies per nucleon-nucleon collision, 
$\sqrt{S_{NN}}$, and 
luminosities for heavy ion collisions at the LHC.  The $t \overline t$ 
production 
cross sections and rates in a one month ($10^6$ s) $AA$ run are also given.}
\label{constant}
\end{table}

\begin{table}
\begin{tabular}{lcccc}
Ion Species & $\sqrt{S_{NN}}$  & $pA$ Luminosity & $\sigma(pA)$ & Rate \\
            & (TeV) & (pb$^{-1}$s$^{-1}$) & (pb) & (per month) \\
\hline
Oxygen  & 9.9  & $3.8\times10^{-4}$  & 2.9  & 1100  \\
Lead    & 8.8  & $7.3\times10^{-7}$  & 55   &   40  \\
\end{tabular}
\caption[]{Center of mass energies per nucleon-nucleon collision, 
$\sqrt{S_{NN}}$, and 
luminosities for $pA$ interactions at the LHC.  The $t \overline t$ 
production cross sections and rates in a $10^6$ s $pA$ run are also given.}
\label{pa}
\end{table}

\begin{figure}
\setlength{\epsfxsize=0.7\textwidth}
\setlength{\epsfysize=0.4\textheight}
\centerline{\epsffile{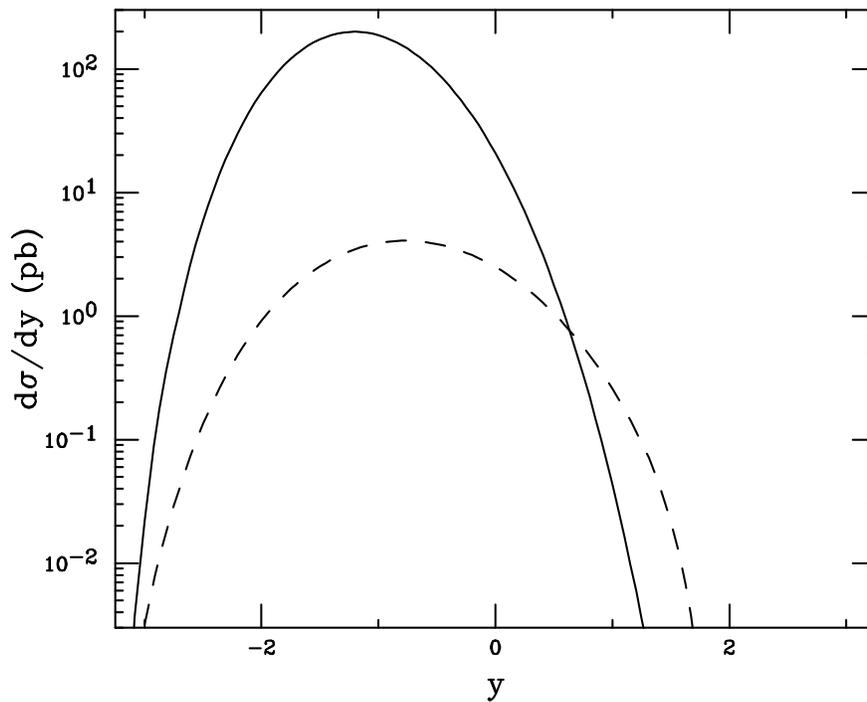}}
\label{dndy}
\caption[]{The top quark rapidity distribution, $d\sigma/dy$, for photoproduced
top at the LHC.
The solid curve is for Pb+Pb collisions while the dashed curve
is for O+O collisions.  In this plot, the photon is emitted by the nucleus
with positive rapidity.}
\end{figure}
\newpage
\begin{figure}
\setlength{\epsfxsize=0.7\textwidth}
\setlength{\epsfysize=0.4\textheight}
\centerline{\epsffile{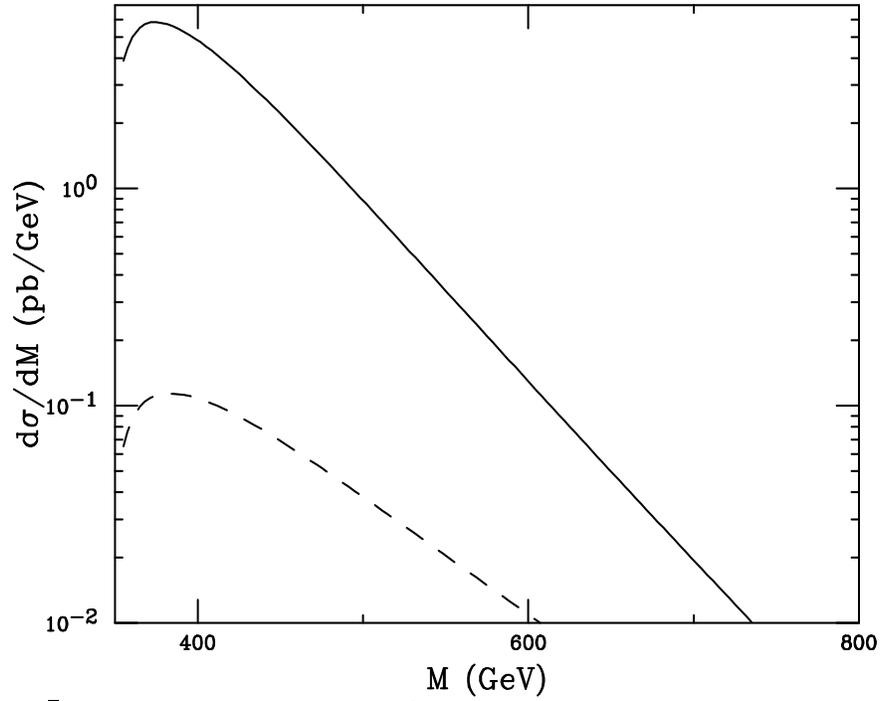}}
\label{dndm}
\caption[]{The $t \overline t$ pair mass distribution, $d\sigma/dM$, for
photoproduced top at
the LHC.  The solid curve shows the results for Pb+Pb collisions while the
dashed curve is for O+O collisions.}
\end{figure}

\end{document}